\documentclass[11pt]{article}

\usepackage{amsmath}
\usepackage{graphicx}
\usepackage{amsfonts}
\usepackage{amssymb}
\usepackage{epsfig}
\usepackage{color}
\usepackage{psfrag}
\usepackage{epstopdf}

\newcommand{\EQ}{\begin{equation}}
\newcommand{\EN}{\end{equation}}
\newcommand{\be}{\begin{equation}}
\newcommand{\ee}{\end{equation}}
\newcommand{\bea}{\begin{eqnarray}}
\newcommand{\eea}{\end{eqnarray}}

\begin{document} \setcounter{page}{0}
\topmargin 0pt
\oddsidemargin 5mm
\renewcommand{\thefootnote}{\arabic{footnote}}
\newpage
\setcounter{page}{0}
\topmargin 0pt
\oddsidemargin 5mm
\renewcommand{\thefootnote}{\arabic{footnote}}
\newpage
\begin{titlepage}
\begin{flushright}
SISSA 18/2012/EP \\
\end{flushright}
\vspace{0.5cm}
\begin{center}
{\large {\bf Phase separation and interface structure in two
dimensions from field theory}}\\
\vspace{1.8cm}
{\large Gesualdo Delfino and Jacopo Viti}\\
\vspace{0.5cm}
{\em SISSA -- Via Bonomea 265, 34136 Trieste, Italy}\\
{\em INFN sezione di Trieste}\\
\end{center}
\vspace{1.2cm}

\renewcommand{\thefootnote}{\arabic{footnote}}
\setcounter{footnote}{0}

\begin{abstract}
\noindent
We study phase separation in two dimensions in the scaling limit below criticality. The general form of the magnetization profile as the volume goes to infinity is determined exactly within the field theoretical framework which explicitly takes into account the topological nature of the elementary excitations. The result known for the Ising model from its lattice solution is recovered as a particular case. In the asymptotic infrared limit the interface behaves as a simple curve characterized by a gaussian passage probability density. The leading deviation, due to branching, from this behavior is also derived and its coefficient is determined for the Potts model. As a byproduct, for random percolation we obtain the asymptotic density profile of a spanning cluster conditioned to touch only the left half of the boundary.
\end{abstract}
\end{titlepage}

\newpage
\section{Introduction}
Boundary conditions notoriously play an important role in the theory
of phase transitions. For a system of ferromagnetic spins taking
discrete values, a pure phase of type $a$ with
translation invariant spontaneous magnetization below the critical
temperature $T_c$ can be selected fixing all boundary spins to the
value $a$ and then sending the boundary to infinity. On the other
hand, if the spins are fixed to a value $a$ on the left half of the
boundary and to a different value $b$ on the right half, a pattern
of phase separation between phases of type $a$ and type $b$ is
expected in the large volume limit below $T_c$, at least away 
from an interfacial region anchored to the points of the boundary where
boundary conditions change from $a$ to $b$. The properties of the
phase separation and the notion of interface have been extensively
studied both in two and three dimensions through rigorous
\cite{Gallavotti}, exact \cite{Abraham} and approximate \cite{Jasnow}
methods. The most advanced analytic results are available in two
dimensions, where the exact asymptotic magnetization profile has
been obtained for the Ising model \cite{AR,Abraham81} exploiting its lattice
solvability at any temperature. This result shows in particular that
in two dimensions the Ising interface has middle point fluctuations which 
diverge as the square root of the volume, a property previously proved 
for low temperatures in \cite{Gallavotti_72}. The
result for the Ising magnetization profile also admits a simple
interpretation in terms of passage probability of the interface
through a point \cite{FFW}. No exact result for the magnetization profile is available in three dimensions.

In this paper we use field theory as a general framework for the study of 
phase separation in the scaling limit below $T_c$, for
any two-dimensional model possessing a discrete set of ordered
phases and undergoing a continuous phase transition. For a
strip of width $R$ we derive 
the large $R$ asymptotics for the magnetization profiles
along the longitudinal axis in the middle of the strip and show that
a generalization of the Ising result
holds whenever the surface tension between the phases $a$ and
$b$ cannot be decomposed into the sum of smaller surface tensions. 
The formalism explicitly illustrates the role played by the topological 
nature of the elementary 
excitations (domain walls), which for a discrete set of ground states is 
peculiar of the two-dimensional case. 
The interpretation in terms of passage probability
holds in general, with subsequent terms in the large $R$ expansion
accounting for the emergence of an interfacial region with finite width in a 
way that can be understood through renormalization group considerations.

The trajectories on the plane of the domain wall excitations of the field theory are naturally interpreted as the continuum limit of the boundaries of clusters made of nearest neighbors with the same value of the spin. In the last years the scaling properties of cluster boundaries have been extensively studied at criticality in the framework of Schramm-Loewner evolution (SLE, see e.g. \cite{Cardy_SLE} for a review); the application of SLE methods to the off-critical case, on the other hand, is up to now much more limited (see \cite{BBC,MS}). The renormalization group interpretation of our results below $T_c$ is that the cluster boundary connecting the two boundary changing points on the edges and the interfacial curve between the two phases coincide as $R\to\infty$, and then have the same gaussian passage probability density; when $R$ decreases, the interfacial region with finite width emerges via branching of the interface and formation of intermediate clusters. We write down the leading term associated to branching and determine its coefficient for the $q$-state Potts model. When the surface tension between the phases $a$ and $b$ is decomposable the formalism leads to a multiple interface description, the Ashkin-Teller model providing an interesting example of this type.

The paper is organized as follows. In the next section we introduce the 
field theoretical formalism and derive the large $R$ results for the 
magnetization profiles. Section~3 is then devoted to the interpretation 
of the results and to the discussion of
the interface structure. The specific cases of $q$-state Potts and
Ashkin-Teller models as well as an application to percolation are
finally discussed in section~4.

\section{Field theoretical results}
Consider a ferromagnetic spin model of two-dimensional classical statistical
mechanics in which each spin can take $n$ discrete values that we
label by an integer $a=1,2,\ldots,n$. To be definite we refer to the
case in which the energy of the system is invariant under global
transformations of the spins according to a symmetry group; the
spontaneous breaking of the symmetry below a critical temperature
$T_c$ is responsible for the presence on the infinite plane of $n$ translation invariant pure phases; the phase of type $a$ can be selected starting with the system on a finite domain with boundary spins fixed to the value $a$, and then removing the boundary to infinity. We denote by $Z_{a}$ and
$\langle\cdots\rangle_{a}$ the partition function and the
statistical averages, respectively, in the phase $a$.

We consider the scaling limit below $T_c$, described by a Euclidean
field theory defined on the plane with coordinates $(x,y)$;
it corresponds to the analytic continuation to imaginary time of a
(1+1)-dimensional relativistic field theory with space coordinate
$x$ and time coordinate $t=iy$. This theory possesses degenerate
vacua $|\Omega_a\rangle$ associated to the 
pure phases of the system. In 1+1 dimensions the elementary
excitations will correspond to stable kink states
$|K_{ab}(\theta)\rangle$ which interpolate between different vacua\footnote{In a kink state $|K_{ab}(\theta)\rangle$ the spin field $\sigma(x,y)$ takes the value $\langle\sigma\rangle_a$ as $x\to-\infty$ and the value $\langle\sigma\rangle_b$ as $x\to+\infty$. Propagation of the kink in Euclidean time $y=it$ corresponds to domain wall configurations for the lattice model.}
$|\Omega_a\rangle$ and $|\Omega_b\rangle$; the rapidity $\theta$
parameterizes the energy and momentum of the kinks as
$(e,p)=(m_{ab}\cosh\theta,m_{ab}\sinh\theta)$, $m_{ab}$ being the
kink mass. In general, connecting $|\Omega_a\rangle$ and $|\Omega_b\rangle$ requires a
multi-kink state $|K_{aa_1}(\theta_1)K_{a_1a_2}(\theta_2)\ldots K_{a_{n-1}b}(\theta_n)\rangle$ passing through other vacua; we call adjacent vacua
two vacua that can be connected through a single-kink excitation. There can be kinks with different masses connecting  two adjacent vacua $|\Omega_a\rangle$ and $|\Omega_b\rangle$; in such a case the notations $|K_{ab}(\theta)\rangle$ and $m_{ab}$ refer to the kink with the lowest mass, which is leading in the large distance limits we will consider.


\begin{figure}[t]
\begin{center}
\includegraphics[width=10cm]{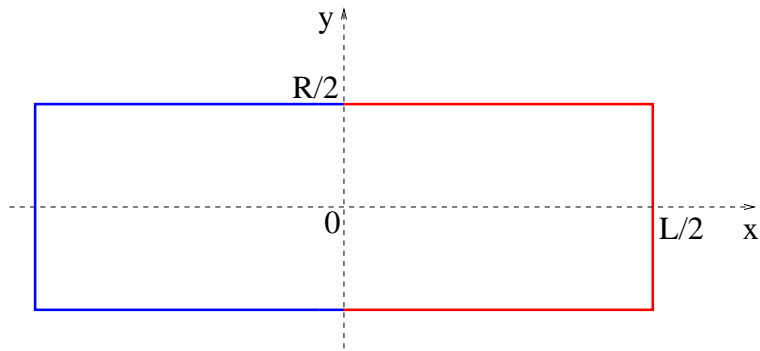}
\caption{The scaling limit below $T_c$ of a ferromagnet is considered on a rectangle with boundary spins fixed to take the value $a$ for $x<0$ and a different value $b$ for $x>0$. The case $L\to\infty$ and $R\gg\xi$ is considered throughout the paper.}
\label{geometry}
\end{center}
\end{figure}

Consider now the scaling limit on a rectangle (Fig.~\ref{geometry}) 
with horizontal sides of
length $L$ and vertical sides of length $R$ (the origin of
the coordinates is taken in the center of the rectangle), with the following choice of boundary conditions (boundary conditions of type $ab$): the boundary spins are fixed to a value $a$ for $x<0$, and to a
different value $b$ for $x>0$. Let us denote by $Z_{ab}$ 
and $\langle\cdots\rangle_{ab}$ the partition function and the
statistical averages  for the system with this choice
of boundary conditions. We consider the limit $L\to\infty$ and want
to study properties of the system as a function of the width $R$ of
the resulting infinite strip, focusing on the asymptotic limit in
which $R$ is much larger than the correlation length $\xi$.

Within the field theoretical formalism the boundary condition at time $t$ switching from $a$ to $b$ at a point $x_0$ is realized by a boundary state  that we denote by $|B_{ab}(x_0;t)\rangle$. This state can be expanded over the basis of asymptotic particle states of the relativistic theory. The
change of boundary conditions at  the point ${x_0}$
requires that kink excitations interpolating between $|\Omega_a\rangle$
and $|\Omega_b\rangle$ are emitted/absorbed at that point.
Then, if $|\Omega_a\rangle$ and $|\Omega_b\rangle$ are adjacent
vacua, the boundary state has the form 
\EQ
|B_{ab}(x_0;t)\rangle=e^{-itH+ix_0P}\left[\int_{-\infty}^{\infty}~\frac{d\theta}{2\pi}f_{ab}(\theta)|
K_{ab}(\theta)\rangle + \ldots\right] \,,
\label{Bab}\EN
where $H$ and $P$ are the energy and momentum operators of the (1+1)-dimensional
theory, $f_{ab}(\theta)$ is the amplitude\footnote{We use the normalization $\langle
K_{ab}(\theta)|K_{a'b'}(\theta')\rangle=2\pi\delta_{aa'}\delta_{bb'}\delta(\theta-\theta')$.} for the kink to be emitted at the boundary changing point, and the dots correspond to states with total mass larger than $m_{ab}$.
The partition function we are considering can be
written as 
\EQ 
Z_{ab}(R)=\langle B_{ab}(x_0;iR/2)|B_{ab}(x_0;-iR/2)\rangle=\langle
B_{ab}(0;0)|e^{-RH}|B_{ab}(0;0)\rangle\,
\EN 
and, as a consequence of (\ref{Bab}), has the large $R$ asymptotics
\begin{equation}
\label{pfab}
Z_{ab}(R)\sim\int_{-\infty}^{\infty}\frac{d\theta}{2\pi}~{e}^{-m_{ab}R\cosh\theta}|f_{ab}(\theta)|^2
\sim\frac{|f_{ab}(0)|^2}{\sqrt{2\pi m_{ab}R}}\,{e}^{-m_{ab}R}.
\end{equation}
The specific interfacial free energy, or surface tension, is given by
\begin{equation}
\Sigma_{ab}=-\lim_{R\rightarrow\infty}\frac{1}{R}\,\ln\frac{Z_{ab}(R)}{Z_{a}(R)}\,,
\end{equation}
where $Z_a(R)$ is the partition function for uniform boundary conditions of type $a$ on the strip.
Since the lowest mass state entering the expansion of the boundary state $|B_a(t)\rangle$ for uniform boundary condition is the vacuum $|\Omega_a\rangle$, $Z_a(R)$ tends to $\langle\Omega_a|\Omega_a\rangle=1$ as $R\to\infty$, so that 
\EQ
\Sigma_{ab}=m_{ab}\,.
\EN 
If $|\Omega_a\rangle$ and $|\Omega_b\rangle$ are not adjacent vacua the expansion of the boundary state $|B_{ab}(x_0;t)\rangle$ starts with a multi-kink state, and the corresponding surface tension is a sum of surface tensions between adjacent vacua; we defer to section~4 the illustration of this case.

Let us denote by $\sigma$ a generic component of the spin field, omitting for the time being the index which in general labels the different components.
The magnetization profile along the horizontal axis in the middle of the strip for $ab$ boundary conditions and $a$ and $b$ adjacent phases is 
\bea
\label{sigmaab}
&& \langle\sigma(x,0)\rangle_{ab} = \frac{1}{Z_{ab}}\langle
B_{ab}(0;0)|\text{e}^{-HR/2+ixP}\sigma(0,0)\text{e}^{-HR/2-ixP}|B_{ab}(0;0)\rangle\\
&& \sim \frac{1}{Z_{ab}}\int\frac{d\theta}{2\pi}\frac{d\theta'}{2\pi}f^*(\theta)f(\theta')\langle
K_{ab}(\theta)|\sigma(0,0)|K_{ab}(\theta')\rangle
e^{m_{ab}[-(\cosh\theta+\cosh\theta')R/2+i(\sinh\theta-\sinh\theta')x]}\,,\nonumber
\eea
the last line being the large $R$ limit obtained from (\ref{Bab}). The matrix element of the spin field it contains is related by the crossing relation\footnote{In field theory `crossing' a particle from the initial to the final state (or vice versa) involves reversing the sign of its energy and momentum. Within the parameterization introduced above this amounts to a $i\pi$ rapidity shift.} 
\begin{equation}
\label{crossing} 
\langle
K_{ab}(\theta)|\sigma(0,0)|K_{ab}(\theta')\rangle=F^{\sigma}_{aba}(\theta-\theta'+i\pi)+2\pi\delta(\theta-\theta')\langle\sigma\rangle_{a}\,,
\end{equation}
to the form factor
\EQ
F^{\sigma}_{aba}(\theta_1-\theta_2)\equiv\langle\Omega_a|\sigma(0,0)|K_{ab}(\theta_1)K_{ba}(\theta_2)\rangle\,,
\label{ff}
\EN
where the vacuum expectation value $\langle\sigma\rangle_a$ appearing in the disconnected part is the spontaneous magnetization in the phase $a$ on the infinite plane.
When $\theta_1-\theta_2=i\pi$ the kink and the anti-kink in (\ref{ff}) have opposite energy and momentum and can annihilate each other. In 1+1 dimensions these annihilation configurations produce in general simple poles that have been characterized for general $k$-particle form factors in integrable field theories (see in particular \cite{F.Smirnov}). For $k=2$, however, integrability plays no role in the determination of the residue, which for the case of kink excitations reads \cite{DC98}
\begin{equation}
\text{Res}_{\theta=i\pi}F_{aba}^{\sigma}(\theta)=i\bigl[\langle\sigma\rangle_{a}-\langle\sigma\rangle_{b}\bigr].
\label{residue}
\end{equation}
For $R\to\infty$ the integral in (\ref{sigmaab}) is dominated by small rapidities and the leading contribution can be written as
\begin{equation}
\label{saddle}
\langle\sigma(x,0)\rangle_{ab}\sim\langle\sigma\rangle_{a}+~\frac{i}{2\pi}\bigl[\langle\sigma\rangle_a-\langle\sigma\rangle_b\bigr]~
\int_{-\infty}^{\infty}d\theta_{-}\frac{1}{\theta_-}\,{e}^{-m_{ab}R\theta_-^2/8+im_{ab}x\theta_-}\,,
\end{equation}
where we used (\ref{pfab}),  (\ref{crossing}) and (\ref{residue}), $\theta_-\equiv\theta-\theta'$, and we integrated over $\theta_{+}\equiv\theta+\theta'$. The last integral is regularized moving the pole slightly above the real axis, so that the usual relation $(x-i0)^{-1}=i\pi\delta(x)+\mbox{p.v.}\,x^{-1}$ finally gives
\EQ
\langle\sigma(x,0)\rangle_{ab}\sim\frac{1}{2}\bigl[\langle\sigma\rangle_{a}+\langle\sigma\rangle_b\bigr]-\frac{1}{2}\bigl[\langle\sigma\rangle_a-\langle\sigma\rangle_b\bigr]\text{erf}\Bigl(\sqrt{\frac{2m_{ab}}{R}}\,x\Bigr)\,,
\label{erfc}
\EN
where the principal value of the integral in (\ref{saddle}) has been expressed in terms of the error function $\text{erf}(x)\equiv\frac{2}{\sqrt{\pi}}\int_{0}^x d\eta~\text{e}^{-\eta^2}$ (see e.g. \cite{Lebedev}). The same result can be obtained differentiating (\ref{saddle}) with respect to $x$ in order to get rid of the pole, and then integrating the result of the integral over $\theta_-$ with the condition $\langle\sigma(+\infty,0)\rangle_{ab}=\langle\sigma\rangle_b$.

For the Ising model ($\langle\sigma\rangle_a=-\langle\sigma\rangle_b$) the result (\ref{erfc}) coincides with the scaling limit of that obtained from the lattice in \cite{AR,Abraham81}. Even in our more general setting, it shows that for $R\to\infty$ $\langle\sigma(\alpha(m_{ab}R)^\beta/m_{ab},0)\rangle_{ab}$ tends to the pure values $\langle\sigma\rangle_a$ or $\langle\sigma\rangle_b$ for $\beta>1/2$ and $\alpha$ negative or positive, respectively, and to the average value $(\langle\sigma\rangle_{a}+\langle\sigma\rangle_b)/2$ for $\beta<1/2$.

The result (\ref{erfc}) is produced by the leading term in the small rapidity expansion of the emission amplitude in (\ref{Bab}) and of the matrix element (\ref{crossing}). More generally, for the latter we write 
\EQ
F_{aba}^\sigma(\theta+i\pi)=\sum_{k=-1}^{\infty}c_{ab}^{(k)}\,\theta^k\,,
\label{ffexpansion}
\EN
with $c_{ab}^{(-1)}$ given by (\ref{residue}). As for the emission amplitude, it satisfies $f_{ab}(\theta)=f_{ba}(-\theta)$ as a consequence of reflection symmetry about the vertical axis. In any model in which $a$ and $b$ play a symmetric role we will have $f_{ab}(\theta)=f_{ba}(\theta)$ and $f_{ab}(\theta)=f_{ab}(0)+O(\theta^2)$. Then it is easy to check that the next contribution to $\langle\sigma(x,0)\rangle_{ab}$ produced by the small rapidity expansion is
\EQ
c_{ab}^{(0)}\sqrt{\frac{2}{\pi m_{ab}R}}\,e^{-2m_{ab}x^2/R}\,.
\label{c0}
\EN
Notice that the error function in (\ref{erfc}) is leading with respect to (\ref{c0}) as $R\to\infty$ for $x\sim (m_{ab}R)^\beta/m_{ab}$ with $\beta>0$; the two terms are of the same order for $\beta=0$.

\section{Passage probability and interface structure}
\begin{figure}[t]
\begin{center}
\includegraphics[width=9cm]{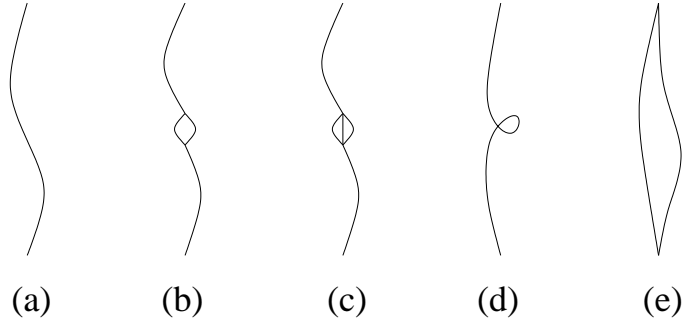}
\caption{Some configurations of the interfacial region. Lines correspond to domain walls between different phases.}
\label{interface}
\end{center}
\end{figure}

The results of the previous section allow an interpretation based on 
renormalization group and probabilistic
considerations. It is clear that the problem has two length scales:
the correlation length $\xi$, proportional to the inverse of the kink masses,  
which is the scale of the fluctuations
within the pure phases, and the width $R$ of the strip, which sets the
scale at which we observe the system with $ab$ boundary conditions. 
In an expansion around
$R/\xi=\infty$ the leading term corresponds to the crudest
description of phase separation in which all short distance features
are washed out and one is left with two pure phases sharply separated by a
simple curve connecting the two boundary changing points (Fig.~\ref{interface}a). Hence, the notion
of curvilinear interface naturally arises in this limit and can be 
formulated directly in the continuum. It is clear, however, that this
picture cannot hold true for finite values of $R/\xi$, and that one needs to
switch from the notion of sharp separation and curvilinear interface to that 
of an interfacial region (or thick interface) with a width which diverges as the correlation length when the critical point is approached: such a divergence simply reflects the fact that phase separation disappears at criticality. Within the large $R/\xi$ expansion the leading deviations from the 
simple curvilinear picture are expected from effects such as branching and
recombination as well as self-intersection of the curve\footnote{At a later stage in the expansion thickness is generated also by 
multi-kink terms in the boundary state (\ref{Bab}), which 
will produce a bundle of thin interfaces rather than just one 
(Fig.~\ref{interface}e). Multi-kink
contributions to $Z_{ab}$ are suppressed at large $R$ as $e^{-MR}$, $M$ being 
the total mass.} (Fig.~\ref{interface}b-d). These effects appear in the expansion through insertions (delta functions)
localized on the curve separating the two pure phases.

According to this discussion the large $R/\xi$ expansion for the
magnetization at a point $x$ on the axis $y=0$ when the interface
passes through a point $u$ on this axis can be expected to start as
\begin{equation}
\sigma_{ab}(x|u)=\theta(u-x)\langle\sigma\rangle_a+
\theta(x-u)\langle\sigma\rangle_b+A_{ab}^{(0)}\delta(x-u)+
A_{ab}^{(1)}\delta'(x-u)+\ldots\,,
\label{RG} 
\end{equation}
where $\theta(x)$ is the step function equal to 1 for $x>0$ and 
zero for $x<0$, and the prime denotes 
differentiation. If $p_{ab}(u)du$ is the
probability that the curve intersects the axis $y=0$ in the
interval $[u,u+du]$, with $p_{ab}(u)=p_{ab}(-u)$ and $\int_{-\infty}^{\infty}
du~p_{ab}(u)=1$, then the average magnetization is
\bea
\langle\sigma(x,0)\rangle_{ab} &= &
\int_{-\infty}^{\infty}du\,p_{ab}(u)\,\sigma_{ab}(x|u)+\ldots\nonumber \\
&=& \langle\sigma\rangle_a\int_{x}^{\infty}du~p_{ab}(u)+
\langle\sigma\rangle_b\int^{x}_{-\infty}du~p_{ab}(u)+A_{ab}^{(0)}p_{ab}(x)
-A_{ab}^{(1)}p_{ab}'(x)+\ldots, 
\label{probability_dec}
\eea
where the dots in the first line stay for the contribution of multi-kink states. Comparison with
(\ref{erfc}) and (\ref{c0}), as it is easily seen, shows correspondence 
between the two expansions and determines
\begin{equation}
p_{ab}(u)=\sqrt{\frac{2m_{ab}}{\pi R}}~\text{e}^{-2m_{ab}u^2/R}\,,
\label{gaussian_fluct} 
\end{equation}
\EQ 
A_{ab}^{(0)}=c_{ab}^{(0)}/m_{ab}   \,. 
\label{bifurcation}
\EN
The last as well as additional terms in (\ref{probability_dec}) should be 
compared with those produced in (\ref{sigmaab}) by further expansion 
around $\theta=\theta'=0$. The gaussian form of $p_{ab}(u)$ is model
independent and of course coincides with the scaling limit of that
deduced in \cite{FFW} for the Ising model\footnote{The gaussian probability density (\ref{gaussian_fluct}) indicates an effective brownian behavior of the interface. The convergence of the interface to a brownian bridge for all subcritical temperatures has been proved in \cite{GI} for the Ising model and in \cite{CIV} for the $q$-state Potts model. We thank Y. Velenik for bringing these references to our attention.}.
As shown in the previous section the integral, non-local terms in (\ref{probability_dec}) are entirely due to the pole term in (\ref{ffexpansion}), which in turn is produced by the non-locality of the kinks with respect to the spin field\footnote{If we consider the form factor of the energy density $\varepsilon$, the residue at $i\pi$ is given by (\ref{residue}) with $\sigma$ replaced by $\varepsilon$, and vanishes because the expectation value of $\varepsilon$ is the same in all stable phases. This reflects the fact that even for a domain wall excitation the energy density is spatially localized (on the wall).}. 

For any lattice configuration there is a cluster (let us call it the left cluster) formed by nearest neighboring spins with value (color) $a$ and whose external perimeter is formed by the left half of the boundary of the strip together with a path connecting the two boundary changing points. This path, whose identification may be ambiguous and require some lattice dependent prescription, becomes a simple curve in the continuum limit. A second curve connecting the two boundary changing points completes the perimeter of the cluster of color $b$ anchored to the right half of the boundary (the right cluster). In general the two curves, which can touch but not intersect, enclose other clusters in between them. Among these intermediate clusters, those adjacent to the left (right) cluster have color different from $a$ ($b$). The first few terms in (\ref{RG}) and (\ref{probability_dec}) are compatible with a picture in which a uniform magnetization $\langle\sigma\rangle_a$ ($\langle\sigma\rangle_b$) is assigned to the region enclosed by the perimeter of the left (right) cluster: the two curves become coincident as $R/\xi\to\infty$, with asymptotic passage probability density given by (\ref{gaussian_fluct}); the first deviation from this situation as $R/\xi$ decreases is expected to happen via bifurcation and recombination around a cluster of color $c\neq a,b$ (Fig.~\ref{interface}b), and to be associated to the term containing $A_{ab}^{(0)}$. We will see in the next section that this term is indeed absent in the Ising model, where bifurcation is not allowed\footnote{Splitting into an odd number of paths (Fig.~\ref{interface}c), however, is allowed and encoded by subsequent terms in the expansion.}.

\section{Specific models}
{\bf Potts model.} The lattice Hamiltonian \cite{Wu}
\begin{equation}
H_{Potts}=-J\sum_{\langle
\mathbf{x_1},\mathbf{x_2}\rangle}\delta_{s(\mathbf{x_1}),s(\mathbf{x_2})},
\hspace{1cm}s(\mathbf{x})=1,\ldots,q\,,
\end{equation}
is invariant under global permutations of the values of the spins 
$s(\mathbf{x})$. For $J>0$ in two dimensions the phase transition is continuous for $q\leq 4$ and above $J_c$ there are $q$ degenerate vacua located at the vertices of a hypertetrahedron in the (q-1)-dimensional order parameter space. Kinks with equal masses run along the edges of the hypertetrahedron and all the vacua are adjacent according to the definition given in section~2.
The results we obtained for the magnetization profile apply to each component
$\sigma_c(\mathbf{x})\equiv\delta_{s(\mathbf{x}),c}-1/q$, $c=1,\ldots,q$, of 
the spin field;  taking into account that $\sum_{c=1}^q\sigma_c=0$ and  $\langle\sigma_a\rangle_b=\frac{q\delta_{ab}-1}{q-1}\langle\sigma_a\rangle_a$, one obtains
\bea
\langle\sigma_c(x,0)\rangle_{ab}&=&\frac{\langle\sigma_a\rangle_{a}}{2}\left[\frac{q(\delta_{ca}+\delta_{cb})-2}{q-1}-\frac{2q(\delta_{ca}-\delta_{cb})}{q-1}\int_0^xdu\,p(u)
\right.\nonumber\\
&+&\left.[2-q( \delta_{ca}+\delta_{cb})]\,\frac{B}{m}\,p(x)\right]+\ldots\,,
\label{sigmac}
\eea
where $p(u)$ is (\ref{gaussian_fluct}) with $m_{ab}=m$. Potts field theory is integrable \cite{CZ} and from the known form factors \cite{DC98} we obtain 
\EQ
B=\frac{1}{2\sqrt{3}},~\frac{2}{3\sqrt{3}}
\EN
for $q=3,4$, respectively. For $c\neq a,b$ the integral term in (\ref{sigmac}) is absent and the $x$-dependence of the magnetization profile is entirely due to the structure of the interface. The gaussian term in (\ref{sigmac}) is produced by the leading deviation from the picture of the interface as a simple curve separating the phases $a$ and $b$, which was argued in the previous section to correpond to the appearance of an island of the phase $c$ via bifurcation and recombination of the curve. Bifurcation is not allowed in the Ising model, and indeed the coefficient of $p(x)$ vanishes at $q=2$, where $c$ necessarily coincides with $a$ or $b$. Directly at $q=2$, the same conclusion is obtained from the explicit form $F_{aba}^\sigma(\theta)=i\langle\sigma\rangle_a\tanh(\theta/2)$ of the spin form factor, implying that (\ref{ffexpansion}) contains only the terms with $k$ odd.

\begin{figure}[t]
\begin{center}
\includegraphics[width=9cm]{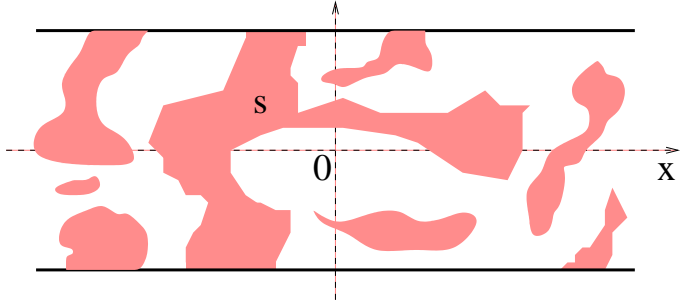}
\caption{For percolation on the strip of width $R$, consider only the configurations without clusters connecting the part of the edges with $x<0$ to the part with $x>0$. Eq.~(\ref{percolation}) gives the probability that a point on the $x$ axis belongs to a cluster $s$ spanning between the negative part of the edges, in the scaling limit above $p_c$ and for $R\gg\xi=1/2m$.}
\label{conditioned}
\end{center}
\end{figure}

\vspace{.2cm}
\noindent
{\bf Percolation.} It is well known \cite{FK,Wu} that the partition function of the Potts model admits an expansion over Fortuin-Kasteleyn (FK) clusters\footnote{The FK clusters differ from the 'geometrical' spin clusters we referred to in the previous section for the fact that nearest neighbors with the same color do not necessarily belong to the same cluster.} of spins with the same color which as $q\to 1$ become the clusters of random percolation. If we consider the Potts model with boundary conditions of type $ab$ on the strip, the probability $\langle\delta_{s(x,0),a}\rangle_{ab}$ that the spin $s(x,0)$ has color $a$ is given by the probability $P(x,0)$ that it belongs to a cluster touching the part of the boundary with $x<0$ (which has color $a$), plus $1/q$ times the probability $1-P(x,0)-P(-x,0)$ that it belongs to a bulk cluster. 
This can be rewritten as
\begin{equation}
\langle\sigma_a(x,0)\rangle_{ab}=\frac{q-1}{q}P(x,0)-\frac{1}{q}P(-x,0)\,.
\label{clusters}
\end{equation}
The FK expansion of $Z_{ab}$ does not contain configurations with
clusters connecting the boundary regions with $x<0$ and $x>0$; this restriction is inherited by the percolation problem we consider. When the occupation probability $p$ for the sites is above the percolation threshold $p_c$ (this corresponds to our case $J>J_c$ in the Potts model), even for $R\to\infty$ there is a positive probability of having a spanning cluster which connects the upper and lower parts of the boundary with $x<0$ (Fig.~\ref{conditioned}). Then the probability $P(x,0)|_{q=1}$ that the site $(x,0)$ is connected to the left part of the boundary is given by the probability $P_s(x,0)$ that it belongs to such spanning cluster plus the probability $P_{ns}(x,0)$ that it belongs to a cluster touching only the upper or lower edge. Since the clusters of the latter type have an average linear extension of order $1/m$, $P_{ns}(x,0)$ vanishes exponentially when $mR\to\infty$. Hence in this limit we have
$P_s(x,0)\sim P(x,0)|_{q=1}$, and from (\ref{sigmac}), (\ref{clusters}) we obtain
\begin{equation}
P_s(x,0)=\frac{P}{2}\left[1-\text{erf}\Bigl(\sqrt{\frac{2m}{R}}\,x\Bigr)-
\gamma\sqrt{\frac{2}{\pi mR}}\,e^{-2mx^2/R}+\ldots\right]\,,
\label{percolation}
\end{equation}
where $P=\lim_{q\to 1}\frac{q}{q-1}\,\langle\sigma_a\rangle_a$
is the probability that a site belongs to the infinite cluster on the infinite plane, and the mass $m$ is related to the exponential correlation length above $p_c$ as $\xi=1/2m$ (see \cite{DVC,crossing}); $\gamma\equiv\text{Res}_{q=1}B(q)$ is not known because for $q\to 1$ the form factor (\ref{ff}) is available only for real rapidities \cite{DVC}. 

\vspace{.2cm}
\noindent
{\bf Ashkin-Teller model.} The lattice model corresponds to two Ising spins $\sigma_1({\bf x}),\sigma_2({\bf x})=\pm 1$ on each site interacting as specified by the Hamiltonian
\EQ
H_{AT}=-\sum_{\langle{\bf x_1,x_2}\rangle}\left\{J[\sigma_1({\bf x_1})\sigma_1({\bf x_2})+
\sigma_2({\bf x_1})\sigma_2({\bf x_2})]+J_4\,\sigma_1({\bf x_1})\sigma_1({\bf x_2})\sigma_2({\bf x_1})\sigma_2({\bf x_2})\right\}\,.
\EN
Each site can be in one of four states $(\sigma_1,\sigma_2)$ that we label $a=1,2,3,4$, corresponding to $(+,+)$, $(+,-)$, $(-,-)$, $(-,+)$, respectively.
The model, that we consider for $J>0$, is well known to possess a line of critical points parameterized by $J_4$ \cite{Baxter}. In the scaling limit close to this line it renormalizes onto the sine-Gordon field theory (see \cite{AT,DG}), where a parameter $\beta^2$ plays the role of $J_4$; $\beta^2=4\pi$ describes a free fermionic theory and corresponds to the decoupling point $J_4=0$. Below critical temperature the model possesses four degenerate vacua $|\Omega_a\rangle$ and for any value of $\beta^2$ there are kinks $|K_{a,a\pm 1(mod\,4)}\rangle$ with the same mass $m$. 

For $\beta^2<4\pi$ the interaction among these kinks (which correspond to sine-Gordon solitons) is attractive and produces bound states, the lightest of which have mass $m'=2m\sin\frac{\pi\beta^2}{2(8\pi-\beta^2)}$ and are kinks $|K_{a,a\pm 2(mod\,4)}\rangle$. Hence, in this regime any pair of vacua is connected by a single-kink excitation, all the vacua are adjacent and the boundary state $|B_{ab}\rangle$ has in any case the form (\ref{Bab}). The results of the previous sections apply with surface tensions $\Sigma_{a,a\pm 1(mod\,4)}=m$ and $\Sigma_{a,a\pm 2(mod\,4)}=m'$; the bifurcation coefficients (\ref{bifurcation}) can be obtained from the form factors computed in \cite{AT,DG}. For $\beta^2=2\pi$ the masses $m$ and $m'$ coincide and one recovers the $q=4$ Potts model.

For $\beta^2\geq 4\pi$ there are no bound states and the vacua with indices differing by two units are not adjacent, with surface tension $\Sigma_{a,a\pm 2(mod\,4)}=2m$. In this case (\ref{Bab}) is replaced by
\EQ
|B_{a,a\pm 2}(x_0;t)\rangle=e^{-itH+ix_0P}\left[\sum_{c=a\pm 1}\int~\frac{d\theta_1}{2\pi}\frac{d\theta_2}{2\pi}\,f_{a,c,a\pm 2}(\theta_1,\theta_2)\,|K_{ac}(\theta_1)K_{c,a\pm 2}(\theta_2)\rangle + \ldots\right],
\label{Bab2}
\EN
(indices are intended mod\,4) and (\ref{RG}) has to be replaced by a description in terms of two interfaces (Fig.~\ref{interface}e).

Let us mention that studies of cluster boundaries at criticality can be found in particular in \cite{GC,DJS} for the Potts model and in \cite{PS,CLR,IR} for the Ashkin-Teller model; we refer the reader to \cite{FKSZ} for results on cluster densities in critical percolation.

\vspace{.3cm}
In summary, in this paper we have shown how field theory naturally accounts for phase separation in two dimensions in the scaling limit below criticality. The derivation is simple but requires to take into account that the large distance properties are determined by kink (domain wall) excitations. In the last decades two-dimensional field theories with kinks have been studied within the non-lagrangian framework of integrable field theory based on asymptotic states, form factors and spectral sums; the derivation of section~2 exploits this framework but is general and does not require integrability. We have shown how it naturally leads to a gaussian passage probability density for the interface and accounts for the deviations from curvilinear behavior. The leading deviation is produced by branching of the inferface and has been determined exactly for the Potts model. Specializations of the formalism to percolation and to the Ashkin-Teller model have also been discussed.


\vspace{1cm} \noindent \textbf{Acknowledgments.} We thank the
Galileo Galilei Institute for Theoretical Physics for the
hospitality during the final stages of this work.

\end{document}